# Beam optics and lattice design for particle accelerators

*Bernhard J. Holzer*
CERN, Geneva, Switzerland

**Abstract**
The goal of this manuscript is to give an introduction into the design of the magnet lattice and as a consequence into the transverse dynamics of the particles in a synchrotron or storage ring. Starting from the basic principles of how to design the geometry of the ring we will briefly review the transverse motion of the particles and apply this knowledge to study the layout and optimization of the principal elements, namely the lattice cells. The detailed arrangement of the accelerator magnets within the cells is explained and will be used to calculate well defined and predictable beam parameters. The more specific treatment of low beta insertions is included as well as the concept of dispersion suppressors that are an indispensable part of modern collider rings.

## 1  Introduction

Lattice design, in the context in which we shall describe it here, is the design and optimization of the principal elements—the lattice cells—of a (circular) accelerator, and includes the detailed arrangement of the accelerator magnets (for example, their positions in the machine and their strength) used to obtain well-defined and predictable parameters of the stored particle beam. It is therefore closely related to the theory of linear beam optics, which is treated in a number of textbooks and proceedings [1].

### 1.1  Geometry of the ring

Magnetic fields are used in circular accelerators to provide the bending force and to focus the particle beam. In principle, the use of electrostatic fields would be possible as well, but at high momenta (i.e., if the particle velocity is close to the speed of light), magnetic fields are much more efficient. The force acting on the particles, the Lorentz force, is given by

$$\vec{F} = q \cdot (\vec{E} + (\vec{v} \times \vec{B})).$$

Neglecting any electrostatic field, the condition for a circular orbit is given by the equality of the Lorentz force and the centrifugal force:

$$q \cdot v \cdot B = \frac{mv^2}{\rho}.$$

In a constant transverse magnetic field $\vec{B}$, a particle will see a constant deflecting force and the trajectory will be part of a circle, whose bending radius is determined by the particle momentum $p = mv$ and the external $B$ field:

$$B \cdot \rho = p/q.$$

The term $B*\rho$ is called the beam rigidity. Inside a dipole magnet, therefore, the beam trajectory is part of a circle, and the bending angle (sketched in Fig. 1) is

$$\alpha = \frac{\int B \, ds}{B\rho}. \tag{1}$$

For the lattice designer, the integrated $B$ field along the design orbit of the particles (sketched roughly in Fig. 1) is the most important parameter, as it is the value that enters Eq. (1) and defines the field strength and the number of such magnets that are needed for a full circle. By requiring a bending angle of $2\pi$ for a full circle, we obtain a condition for the magnetic dipole fields in the ring. Figure 2 shows a photograph of a small storage ring [2], where only eight dipole magnets are used to define the design orbit. The magnets are powered symmetrically, and therefore each magnet corresponds to a bending angle $\alpha$ of the beam of exactly 45°. The field strength $B$ in this machine is of the order of 1 T.

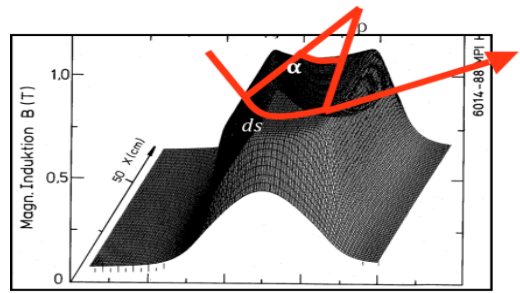

**Fig. 1:** Magnetic $B$ field in a storage ring dipole and, schematically, the particle orbit

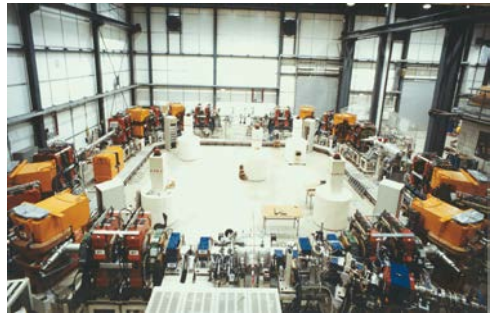

**Fig. 2:** The TSR heavy-ion storage ring at the Max-Planck-Institut in Heidelberg

In the case of the Large Hadron Collider LHC at CERN, for a momentum $p = 7000$ GeV/$c$, 1232 dipole magnets are needed, each having a length of 15 m and a $B$ field of 8.3 T. As a general rule, in high-energy rings, about 66% of the circumference of the machine is used to install dipole magnets and in that way define the maximum momentum (or energy) of particles that can be stored in the machine. The remaining part of the circumference is equipped with focusing elements, RF systems for particle acceleration, diagnostics, and long straight sections for the installation of high-energy detectors.

The lattice and, correspondingly, the beam optics are therefore split into several different characteristic parts. These include arc structures, which are used to guide the particle beam and establish a regular pattern of focusing elements, leading to a regular, periodic $\beta$-function. These structures define the geometry of the ring and, as a function of the installed dipole magnets, the maximum energy of the stored particle beam. The arcs are connected by so-called insertions, long lattice sections where the optics is modified to establish the conditions needed for particle injection, to reduce the dispersion function, or to reduce the beam dimensions in order to increase the particle collision rate (for example, where the beam in a collider ring must be prepared for particle collisions).

## 1.2 Equation of motion and matrix formalism

Once the geometry and specification of the arc have been determined and the layout of the bending magnets has been done, the next step is to worry about the focusing properties of the machine. In general, we have to keep more than $10^{12}$ particles in the machine, distributed over a number of bunches, and these particles have to be focused to keep their trajectories close to the design orbit.

As we have heard in the lecture on linear beam optics, gradient fields generated by quadrupole lenses are used to do this job. These lenses generate a magnetic field that increases linearly as a function of the distance from the magnet centre:

$$B_y = -g \cdot x, \quad B_x = -g \cdot y \ .$$

Here, $x$ and $y$ refer to the horizontal and vertical planes and the parameter $g$ is called the gradient of the magnetic field. It is customary to normalize the magnetic fields to the momentum of the particles. In the case of dipole fields, we obtain from Eq. (1)

$$\alpha = \frac{\int B ds}{B\rho} = \frac{L_{\text{eff}}}{\rho},$$

where $L_{\text{eff}}$ is the so-called effective length of the magnet. The term $1/\rho$ is the bending strength of the dipole. In the same way, the field of the quadrupole lenses is normalized to $B\rho$. The strength $k$ is defined by

$$k = \frac{g}{B \cdot \rho},$$

and the focal length of the quadrupole is given by

$$f = \frac{1}{k \ell_q} \ .$$

The particle trajectories under the influence of the focusing properties of the quadrupole and dipole fields in the ring are described by a differential equation. This equation is derived in its full beauty in [1], so we shall just state here that it is given by the expression

$$x'' + Kx = 0 \ . \tag{2}$$

Here, $x$ describes the horizontal coordinate of the particle with respect to the design orbit; the derivative is taken with respect to the orbit coordinate $s$, as usual in linear beam optics; and the parameter $K$ combines the focusing strength $k$ of the quadrupole and the weak-focusing term $1/\rho^2$ of the dipole field. (Note: a negative value of $k$ means a horizontal focusing magnet.) $K$ is given by

$$K = -k + 1/\rho^2 \ .$$

In the vertical plane, in general, the term $1/\rho^2$ is missing, as in most accelerators (but not all) the design orbit is in the horizontal plane and no vertical bending strength is present. So, in the vertical plane, we have

$$K = k \ .$$

When we are starting to design a magnet lattice, we ought to make as many simplifications as possible at the beginning of the process. Clearly, the exact solution for the particle motion has to be calculated in full detail, and if the beam optics is optimized on a linear basis, higher-order multipole fields and their effect on the beam have to be taken into account. But when we are doing the very first steps, we can make life a little bit easier and ignore terms that are small enough to be neglected.

In many cases, for example, the weak-focusing term $1/\rho^2$ can be neglected, to obtain a rough estimate that makes the formula much shorter and symmetric in the horizontal and vertical planes. Referring to the HERA proton ring as an example, the basic parameters of this machine are listed in Table 1. The weak-focusing contribution in this case, $1/\rho^2 = 2.97 \times 10^{-6} / m^2$, is indeed much smaller than the quadrupole strength $k$. For initial estimates in the context of the lattices of large accelerators, this contribution can in general be neglected.

**Table 1:** Basic parameters of the HERA proton storage ring

| | |
|---|---|
| Circumference $C_0$ | 6335 m |
| Bending radius $\rho$ | 580 m |
| Quadrupole gradient $G$ | 110 T/m |
| Particle momentum $p$ | 920 GeV/$c$ |
| Weak-focusing term $1/\rho^2$ | $2.97 \times 10^{-6} / m^2$ |
| Focusing strength $k$ | $3.3 \times 10^{-3} / m^2$ |

## 1.3 Single-particle trajectories

The differential equation in Eq. (2) describes the transverse motion of a particle with respect to the design orbit. This equation can be solved in a linear approximation, and the solutions for the horizontal and vertical planes are independent of each other.

If the focusing parameter $K$ is constant, which means that we are referring to a place inside a magnet where the field is constant along the orbit, the general solution for the position and angle of the trajectory can be derived as a function of the initial conditions $x_0$ and $x_0'$. In the case of a focusing lens, we obtain

$$x(s) = x_0 * \cos(\sqrt{K} * s) + \frac{x_0'}{\sqrt{K}} * \sin(\sqrt{K} * s),$$
$$x'(s) = -x_0 * \sqrt{K} * \sin(\sqrt{K} * s) + x_0' * \cos(\sqrt{K} * s),$$

or, written in a more convenient matrix form,

$$\begin{pmatrix} x \\ x' \end{pmatrix}_s = M \cdot \begin{pmatrix} x \\ x' \end{pmatrix}_0.$$

The matrix $M$ depends on the properties of the magnet and, for a number of typical lattice elements, we obtain the following:

focusing quadrupole: $\quad M_{QF} = \begin{pmatrix} \cos(\sqrt{K}l) & \frac{1}{\sqrt{K}}\sin(\sqrt{K}l) \\ -\sqrt{K}\sin(\sqrt{K}l) & \cos(\sqrt{K}l) \end{pmatrix},$ (3a)

$$\text{defocusing quadrupole:} \quad M_{QD} = \begin{pmatrix} \cosh(\sqrt{K}l) & \frac{1}{\sqrt{K}}\sinh(\sqrt{K}l) \\ \sqrt{K}\sinh(\sqrt{K}l) & \cosh(\sqrt{K}l) \end{pmatrix}, \quad (3b)$$

$$\text{drift space:} \quad M_{\text{drift}} = \begin{pmatrix} 1 & \ell \\ 0 & 1 \end{pmatrix}. \quad (3c)$$

## 1.4 The Twiss parameters α, β, γ

In the case of periodic conditions in the accelerator, there is another way to describe the particle trajectories that, in many cases, is more convenient than the above-mentioned formalism, which is valid within a single element. It is important to note that in a circular accelerator, the focusing elements are necessarily periodic in the orbit coordinate $s$ after one revolution. Furthermore, storage ring lattices have in most cases an inner periodicity: they often are constructed, at least partly, from sequences in which identical magnetic cells, the lattice cells, are repeated several times in the ring and lead to periodically repeated focusing properties.

In this case, the transfer matrix from the beginning of such a structure to the end can be expressed as a function of the periodic parameters $\alpha, \beta, \gamma, \varphi$:

$$M(s) = \begin{pmatrix} \cos(\varphi) + \alpha_s \sin(\varphi) & \beta_s \sin(\varphi) \\ -\gamma_s \sin(\varphi) & \cos(\varphi) - \alpha_s \sin(\varphi) \end{pmatrix}. \quad (4)$$

The parameters $\alpha$ and $\gamma$ are related to the $\beta$-function by the equations

$$\alpha(s) = -\frac{1}{2}\beta'(s) \quad \text{and} \quad \gamma(s) = \frac{1+\alpha^2(s)}{\beta(s)}.$$

The matrix is clearly a function of the position $s$, as the parameters $\alpha, \beta, \gamma$ depend on $s$. The variable $\varphi$ is called the phase advance of the trajectory and is given by

$$\varphi = \int_s^{s+L} \frac{d\tilde{s}}{\beta(\tilde{s})}.$$

In such a periodic lattice, the relation

$$|\text{trace}(M)| < 2$$

has to be valid for stability of the equation of motion, which sets boundary conditions for the focusing properties of the lattice, as we shall see in a moment.

Given this correlation, the solution for the trajectory of a particle can be expressed as a function of the following new parameters:

$$x(s) = \sqrt{\varepsilon} \cdot \sqrt{\beta(s)} \cdot \cos(\varphi(s) - \delta),$$

$$x'(s) = \frac{-\sqrt{\varepsilon}}{\sqrt{\beta(s)}} \cdot \{\sin(\varphi(s) - \delta) + \alpha(s)\cos(\varphi(s) - \delta)\}$$

The position and angle of the transverse oscillation of a particle at a point $s$ is given by the value of the $\beta$-function at that location, and $\varepsilon$ and $\delta$ are constants of the particular trajectory.

As a last reminder of the linear beam optics, we state that the Twiss parameters at a position $s$ in the lattice are defined by the focusing properties of the complete storage ring. These parameters are transformed from one point to another in the lattice by the elements of the product matrix of the corresponding magnets. Without proof, we state that if the matrix $M$ is given by

$$M(s_1, s_2) = \begin{pmatrix} C & S \\ C' & S' \end{pmatrix}, \tag{5}$$

the transformation rule from point $s_1$ to point $s_2$ in the lattice is given by

$$\begin{pmatrix} \beta \\ \alpha \\ \gamma \end{pmatrix}_{s_2} = \begin{pmatrix} C^2 & -2SC & S^2 \\ -CC' & SC' + S'C & -SS' \\ C'^2 & -2S'C' & S'^2 \end{pmatrix} * \begin{pmatrix} \beta \\ \alpha \\ \gamma \end{pmatrix}_{s_1}. \tag{6}$$

The terms $C$, $S$, etc. correspond to the focusing properties of the matrix. In the case of a single element, for example, they are just the expressions given in Eq. (3).

## 2  Lattice design

An example of a high-energy lattice and the corresponding beam optics is shown in Fig. 3, for the LHC storage ring. In general, such machines are designed on the basis of small elements, called cells, that are repeated many times in the ring. One of the most widespread lattice cells used for this purpose is the so-called FODO cell, a magnet structure consisting alternately of focusing and defocusing quadrupole lenses. Between the focusing magnet elements, the dipole magnets are installed, and any other machine elements such as orbit corrector dipoles, multipole correction coils, and diagnostics elements.

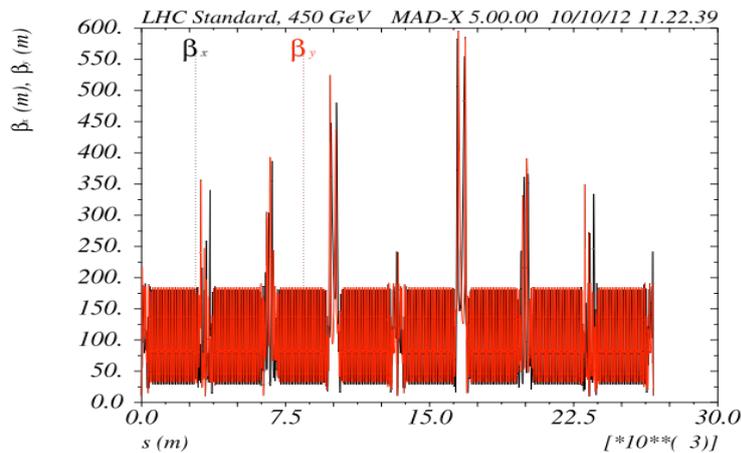

**Fig. 3:** Beam optics of the LHC storage ring

The optical solution for such a FODO cell is plotted in Fig. 4. The graph shows the $\beta$-function in the two transverse planes (solid line for the horizontal plane and dotted line for the vertical plane). The positions of the magnet lenses, i.e., the lattice, are shown schematically in the lower part of the plot. Owing to the symmetry of the cell, the solution for the $\beta$-function is periodic (in general, the FODO is the smallest periodic structure in a storage ring), and it reaches its maximum in the horizontal plane in the focusing lenses and its minimum in the defocusing lenses. Accordingly, the $\alpha$-function is generally zero in the centre of a FODO quadrupole.

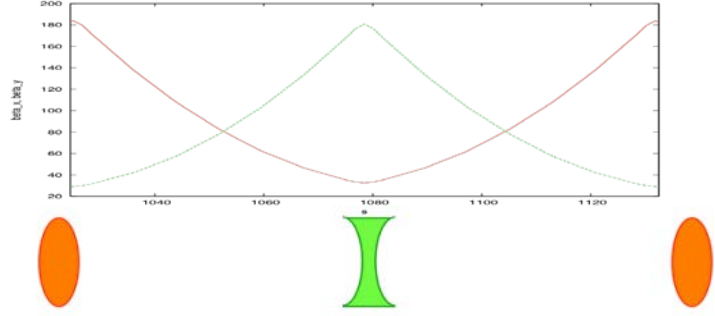

**Fig. 4:** A FODO cell

**Table 2:** Result of an optics calculation for a FODO lattice

| Element | $l$ (m) | $k$ (1/m$^2$) | $\beta_x$ (m) | $\alpha_x$ | $\varphi_x$ (rad) | $\beta_y$ (m) | $\alpha_y$ | $\varphi_y$ (rad) |
|---|---|---|---|---|---|---|---|---|
| Start | 0 | – | 11.611 | 0 | 0 | 5.295 | 0 | 0 |
| QFH | 0.25 | −0.0541 | 11.228 | 1.514 | 0.0110 | 5.488 | −0.78 | 0.0220 |
| QD | 3.251 | 0.0541 | 5.4883 | −0.78 | 0.2196 | 11.23 | 1.514 | 0.2073 |
| QFH | 6.002 | −0.0541 | 11.611 | 0 | 0.3927 | 5.295 | 0 | 0.3927 |
| End | 6.002 | – | 11.611 | 0 | 0.3927 | 5.295 | 0 | 0.3927 |

Table 2 summarizes the main parameters of the lattice magnets (quadrupole gradients, lengths, and positions) and the resulting optical properties of an example of such a single cell, calculated with an optics code. Due to symmetry reasons the calculation starts in the middle of a focusing quadrupole, named QFH in the table. Qualitatively speaking, it is already clear from the schematic drawing in Fig. 5 that the horizontal function $\beta_x$ reaches its maximum value at the centre of the (horizontal) focusing quadrupoles and its minimum value at the defocusing lenses. For the vertical function $\beta_y$, a similar statement holds - vice versa - with 'maximum' and 'minimum' interchanged. The $\alpha$-function in the centre of the quadrupole is indeed zero and, as $\alpha(s) = -\beta'(s)/2$, the $\beta$-function is maximum or minimum at that position.

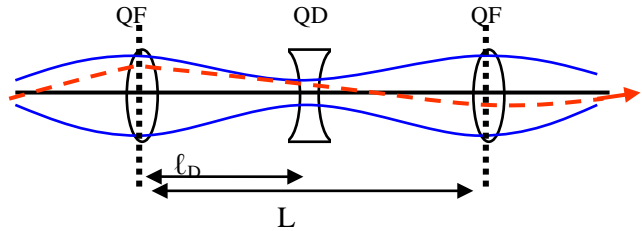

**Fig. 5:** Schematic drawing of a symmetric FODO cell with the beam envelope marked in blue and a single particle trajectory in red.

The phase advance of the complete machine, measured in units of $2\pi$, is called the working point. In our case we have chosen $\varphi = 45°$, which corresponds to 0.3927 rad, as the phase advance of a single cell, and the corresponding working point is $Q_x = \int \varphi \, ds/2\pi = 0.125$. As we have chosen equal quadrupole strengths in the two planes, i.e., $k_x = -k_y$, and uniform drift spaces between the quadrupoles, the lattice is called a symmetric FODO cell. We therefore expect symmetric optical solutions in the two transverse planes.

The question now is: Can we understand what the optics code is doing? For this purpose, we refer to a single cell. In linear beam optics, the transfer matrix of a number of optical elements is given by the product of the matrices of the individual elements. In our case we obtain

$$M_{FODO} = M_{QFH} \cdot M_{Ld} \cdot M_{QD} \cdot M_{Ld} \cdot M_{QFH}. \tag{7}$$

It has to be pointed out that as we have decided to start the calculation in the centre of a quadrupole magnet, the corresponding matrix has to take this into account: the first matrix has to be that of a half quadrupole, QFH. Putting in the numbers for the length and strength $k = \pm 0.54102\ /m^2$, $l_q = 0.5$ m, $l_d = 2.5$ m, where $l_q$ and $l_d$ refer to the length of the quadruple magnets and the drift space between them, we obtain

$$M_{FODO} = \begin{pmatrix} 0.707 & 8.206 \\ -0.061 & 0.707 \end{pmatrix}.$$

As we shall now see, this matrix describes uniquely the optical properties of the lattice and defines the beam parameters.

### 2.1 The most important point: stability of the motion

Taking the trace of $M$ gives

$$|trace(M_{FODO})| = 1.415 < 2.$$

A lattice built out of such FODO cells would therefore give stable conditions for the particle motion. However, if new parts of the lattice are introduced, we have to go through the calculation again, as we shall see later. In addition, the matrix can be used to determine the optical parameters of the system, as described in what follows.

### 2.2 Phase advance per cell

Writing $M$ as a function of $\alpha$, $\beta$, $\gamma$, and the phase advance $\varphi$, we obtain for a periodic situation

$$M(s) = \begin{pmatrix} \cos(\varphi) + \alpha_s \sin(\varphi) & \beta_s \sin(\varphi) \\ -\gamma_s \sin(\varphi) & \cos(\varphi) - \alpha_s \sin(\varphi) \end{pmatrix} \tag{8}$$

and we immediately see that

$$\cos(\varphi) = \frac{1}{2} \cdot trace(M) = 0.707,$$

or $\varphi = 45°$, which corresponds to the working point of 0.125 calculated above.

### 2.3 Calculation of $\alpha$- and $\beta$-functions

The $\alpha$- and $\beta$-functions are calculated in a similar way. For $\beta$, we use the relation

$$\beta = \frac{M(1,2)}{\sin(\varphi)} = 11.611\ \text{m},$$

and we obtain $\alpha$ from the expression

$$\alpha = \frac{M(1,1) - \cos(\varphi)}{\sin(\varphi)} = 0.$$

We can see that these analytical calculations lead to exactly the same results as the optics code used in Table 2.

To complete this first look at the optical properties of a lattice cell, I want to give a rule of thumb for the working point. Defining an average $\beta$-function for the ring, we put

$$\oint \frac{ds}{\beta(s)} \approx \frac{L}{\overline{\beta}}.$$

If we set $L = 2\pi\overline{R}$, where $\overline{R}$ is the geometric radius of the ring (which is *not* the bending radius of the dipole magnets), we can write the following for the working point $Q$:

$$Q = N \cdot \frac{\varphi_c}{2\pi} = \frac{1}{2\pi} \cdot \oint \frac{ds}{\beta(s)} \approx \frac{1}{2\pi} \cdot \frac{2\pi\overline{R}}{\overline{\beta}},$$

where $N$ is the number of cells and $\varphi_c$ denotes the phase advance per cell. So, we obtain

$$Q = \frac{\overline{R}}{\overline{\beta}}.\qquad(9)$$

Therefore a rough estimate of the working point can be obtained from the ratio of the mean radius of the ring to the average $\beta$-function of the lattice.

### 2.4 Thin-lens approximation

As we have seen, an initial estimate of the parameters of a lattice can and should be made at the beginning of the design of a magnet lattice. If we want fast answers and require only rough estimates, we fortunately can make the task a little easier: under certain circumstances, the matrix of a focusing element can be written in the so-called thin-lens approximation.

Given for example the matrix of a focusing lens

$$M_{QF} = \begin{pmatrix} \cos(\sqrt{K}l) & \frac{1}{\sqrt{K}}\sin(\sqrt{K}l) \\ -\sqrt{K}\sin(\sqrt{K}l) & \cos(\sqrt{K}l) \end{pmatrix},$$

we can simplify the trigonometric terms if the focal length of the quadrupole magnet is much larger than the length of the lens: if

$$f = \frac{1}{k\ell_Q} \gg \ell_Q,$$

then the transfer matrix can be approximated using $kl_Q = $ const, $l_Q \to 0$, and we obtain

$$M_{QF} = \begin{pmatrix} 1 & 0 \\ \frac{-1}{f} & 1 \end{pmatrix}$$

Referring to the notation used in Fig. 5, we can calculate first the transfer matrix from the centre of the focusing quadrupole to the centre of the defocusing quadrupole and obtain the matrix for half of the cell:

$$M_{\text{half cell}} = M_{QD/2} M_{\ell_D} M_{QF/2},$$

$$M_{\text{half cell}} = \begin{pmatrix} 1 & 0 \\ \frac{1}{\tilde{f}} & 1 \end{pmatrix} \cdot \begin{pmatrix} 1 & \ell_D \\ 0 & 1 \end{pmatrix} \cdot \begin{pmatrix} 1 & 0 \\ \frac{-1}{\tilde{f}} & 1 \end{pmatrix},$$

$$M_{\text{half cell}} = \begin{pmatrix} 1 - \frac{\ell_D}{\tilde{f}} & \ell_D \\ -\frac{\ell_D}{\tilde{f}^2} & 1 + \frac{\ell_D}{\tilde{f}} \end{pmatrix}. \tag{10}$$

Note that the thin-lens approximation implies that $\ell_Q \to 0$, and therefore the drift length $\ell_D$ between the magnets has to be equal to $L/2$. And, as we are now dealing with half quadrupoles, we must set $\tilde{f} = 2f$ for the focal length of a half quadrupole.

We obtain the second half of the cell simply by replacing $\tilde{f}$ by $-\tilde{f}$, and the matrix for the complete FODO in the thin-lens approximation is

$$M_{FODO} = \begin{pmatrix} 1 + \frac{l_D}{\tilde{f}} & l_D \\ \frac{-l_D}{\tilde{f}^2} & 1 - \frac{l_D}{\tilde{f}} \end{pmatrix} \cdot \begin{pmatrix} 1 - \frac{l_D}{\tilde{f}} & l_D \\ \frac{-l_D}{\tilde{f}^2} & 1 + \frac{l_D}{\tilde{f}} \end{pmatrix}$$

or, multiplying out,

$$M = \begin{pmatrix} 1 - \frac{2\ell_D^2}{\tilde{f}^2} & 2\ell_D \left(1 + \frac{\ell_D}{\tilde{f}}\right) \\ 2\left(\frac{\ell_D^2}{\tilde{f}^3} - \frac{\ell_D}{\tilde{f}^2}\right) & 1 - 2\frac{\ell_D^2}{\tilde{f}^2} \end{pmatrix}. \tag{11}$$

The matrix is now much easier to handle than the equivalent formulae in Eqs. (3) and (7), and the approximation is, in general, not bad.

Going briefly again through the calculation of the optics parameters, we immediately obtain from Eqs. (8) and (11)

$$\cos(\varphi) = 1 - \frac{2\ell_D^2}{\tilde{f}^2}$$

and, with a little bit of trigonometric gymnastics,

$$1 - 2\sin^2(\varphi/2) = 1 - \frac{2\ell_D^2}{\tilde{f}^2}.$$

We can simplify this expression and obtain

$$\sin(\varphi/2) = \frac{\ell_\mathrm{D}}{\tilde{f}} = \frac{L_\mathrm{cell}}{2\tilde{f}},$$

and finally

$$\sin(\varphi/2) = \frac{L_\mathrm{cell}}{4f}. \tag{12}$$

In the thin-lens approximation, the phase advance of a FODO cell is given by the length of the cell $L_\mathrm{cell}$, and the focal length of the quadrupole magnets $f$.

For the parameters of the example given above, we obtain a phase advance per cell of $\varphi \approx 47.8°$ and, in full analogy to the calculation presented earlier, we calculate $\beta \approx 11.4$ m, which is very close to the result of the exact calculation ($\varphi = 45°, \beta = 11.6$ m).

### 2.4.1 Stability of the motion

In the thin-lens approximation, the condition for stability $|\mathrm{trace}(M)| < 2$ requires that

$$\left| 2 - \frac{4\ell_\mathrm{D}^2}{\tilde{f}^2} \right| < 2,$$

or

$$f > \frac{L_\mathrm{cell}}{4}. \tag{13}$$

We have obtained the important and simple result that for stable motion, the focal length of the quadrupole lenses in the FODO has to be larger than a quarter of the length of the cell.

## 2.5 Scaling the optical parameters of a lattice cell

After the above discussion of stability in a lattice cell and initial estimates and calculations of the optical functions $\alpha$, $\beta$, $\gamma$, and $\varphi$, we shall now concentrate a little more on a detailed analysis of a FODO concerning these parameters.

As we have seen, we can calculate the $\beta$-function that corresponds to the periodic solution — provided that we know the strength and length of the focusing elements in the cell. But can we optimize the solution somehow? In other words, for a given lattice, what would be the ideal magnet strength to obtain the smallest beam dimensions? To answer this question, we shall go back to the transfer matrix for half a FODO cell as indicated in Eq. (10), i.e., the transfer matrix from the centre of a focusing quadrupole to the centre of a defocusing quadrupole (see Fig. (4)).

From linear beam optics, we know that the transfer matrix between two points in a lattice can be expressed not only as a function of the focusing properties of the elements in that section of the ring but also, in an equivalent way, as a function of the optical parameters between the two reference points. We have used this relation already in Eq. (4) for a full turn or for one period in a periodic lattice. The general expression, in the non-periodic case, reads [1]

$$M_{1 \to 2} = \begin{pmatrix} \sqrt{\frac{\beta_2}{\beta_1}}(\cos\Delta\varphi + \alpha_1 \sin\Delta\varphi) & \sqrt{\beta_2 \beta_1} \sin\Delta\varphi \\ \frac{(\alpha_1 - \alpha_2)\cos\Delta\varphi - (1 + \alpha_1\alpha_2)\sin\Delta\varphi}{\sqrt{\beta_2 \beta_1}} & \sqrt{\frac{\beta_1}{\beta_2}}(\cos\Delta\varphi - \alpha_2 \sin\Delta\varphi) \end{pmatrix}. \tag{14}$$

The indices refer to the starting point $s_1$ and the end point $s_2$ in the ring, and $\Delta\varphi$ is the phase advance between these points. It is evident that this matrix can be reduced to the form given in Eq. (4) if the periodic conditions $\beta_1 = \beta_2$, $\alpha_1 = \alpha_2$ are fulfilled.

We know already that $\beta$ reaches its highest value in the centre of the focusing quadrupole and its lowest value in the centre of the defocusing magnet (for the vertical plane, the argument is valid 'vice versa'), and the $\alpha$-functions at these positions are zero. Therefore the transfer matrix for a half cell can be written in the form

$$M = \begin{pmatrix} C & S \\ C' & S' \end{pmatrix} = \begin{pmatrix} \sqrt{\dfrac{\check{\beta}}{\hat{\beta}}}\cos\Delta\varphi & \sqrt{\hat{\beta}\check{\beta}}\sin\Delta\varphi \\ \dfrac{-1}{\sqrt{\hat{\beta}\check{\beta}}}\sin\Delta\varphi & \sqrt{\dfrac{\hat{\beta}}{\check{\beta}}}\cos\Delta\varphi \end{pmatrix}.$$

Using this expression and putting for the matrix elements the terms that we have developed in the thin-lens approximation in Eq. (10), we obtain

$$\frac{\hat{\beta}}{\check{\beta}} = \frac{S'}{C} = \frac{1+\ell_D/\tilde{f}}{1-\ell_D/\tilde{f}} = \frac{1+\sin(\varphi/2)}{1-\sin(\varphi/2)},$$

$$\hat{\beta}\check{\beta} = \frac{-S}{C'} = \tilde{f}^2 = \frac{\ell_D^2}{\sin^2(\varphi/2)},$$

where we have set $\Delta\varphi = \varphi/2$ for the phase advance of half the FODO cell. The two expressions can be combined to calculate the two parameters $\hat{\beta}$ and $\check{\beta}$:

$$\hat{\beta} = \frac{(1+\sin(\varphi/2))L}{\sin\varphi}, \quad \check{\beta} = \frac{(1-\sin(\varphi/2))L}{\sin\varphi}. \tag{15}$$

We obtain the simple result that the maximum (and minimum) value of the $\beta$-function and therefore the maximum dimension of the beam in the cell are determined by the length $L$ and the phase advance $\varphi$ of the complete cell.

Figure 6 shows a three-dimensional picture of a proton bunch for typical conditions in the HERA storage ring. The bunch length is about 30 cm and is determined by the momentum spread and the RF potential [3]. The values of $\hat{\beta}$ and $\check{\beta}$, as determined by the cell characteristics, are typically 80 and 40 m, respectively.

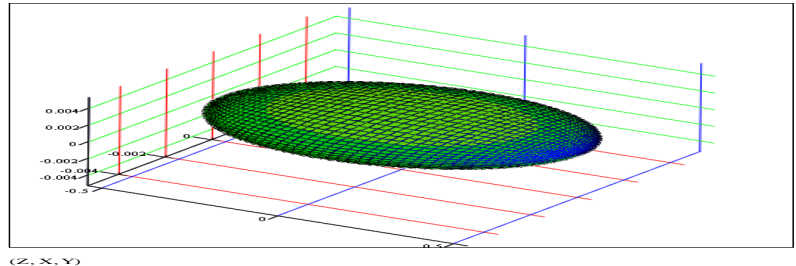

**Fig. 6:** Typical shape of a proton bunch in an arc of the HERA proton ring

### 2.5.1 Optimization of the FODO phase advance

From Eq. (15) we see that—given the length of the FODO—the maximum value of $\beta$ depends only on the phase advance per cell. Therefore we may ask whether there is an optimum phase that leads to the smallest beam dimension.

If we assume a Gaussian particle distribution in the transverse plane and denote the beam emittance by $\varepsilon$, the transverse beam dimension $\sigma$ is given by

$$\sigma = \sqrt{\varepsilon\beta} .$$

In a typical high-energy proton ring, $\varepsilon$ is of the order of some $10^{-9}$ m·rad (e.g., for the HERA proton ring at $E = 920$ GeV, $\varepsilon \approx 6 \times 10^{-9}$ m·rad), and as the typical $\beta$-functions have values of about 40–100 m in an arc, the resulting beam dimension is roughly a millimetre. At the interaction point of two counter-rotating beams, even beam radii of the order of micrometres can be obtained. Figure 7 shows the result of a beam scan that was used to measure the transverse beam dimension.

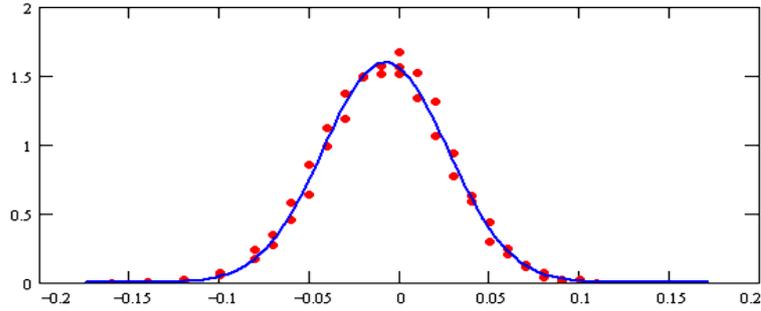

**Fig. 7:** Transverse beam profile of a HERA proton bunch at the interaction point. The measurement was performed by scanning the colliding beams against each other. The data points (dots) have been fitted by a Gaussian curve (line).

In general, the two emittances are equal for a proton beam, i.e., $\varepsilon_x \approx \varepsilon_y$. In this sense, a proton beam is 'round', even if the varying $\beta$-function along the lattice leads to beam dimensions in the two transverse planes that can be quite different. Optimizing the beam dimensions in the case of a proton ring therefore means searching for a minimum of the beam radius given by

$$r^2 = \varepsilon_x \beta_x + \varepsilon_y \beta_y ,$$

and therefore optimizing the sum of the maximum and minimum $\beta$-functions

$$\hat{\beta} + \check{\beta} = \frac{(1+\sin(\varphi/2))L}{\sin\varphi} + \frac{(1-\sin(\varphi/2))L}{\sin\varphi} \qquad (16)$$

at the same time. The optimum phase $\varphi$ is obtained from the condition

$$\frac{\mathrm{d}}{\mathrm{d}\varphi}(\hat{\beta} + \check{\beta}) = \frac{\mathrm{d}}{\mathrm{d}\varphi}\left(\frac{2L}{\sin\varphi}\right) = 0 ,$$

which gives

$$\frac{L}{\sin^2\varphi} * \cos\varphi = 0 \quad \rightarrow \quad \varphi = 90° .$$

Concerning the aperture requirement of the cell, a phase advance of $\varphi = 90°$ is the best value for a proton ring. The plot in Fig. 8 shows the sum of the two $\beta$'s (Eq. 16) as a function of the phase $\varphi$ in the range $\varphi = 0–180°$. The optimization of the beam radii can be a critical issue in accelerator design.

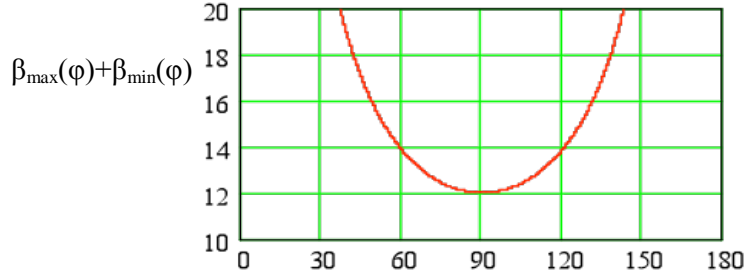

**Fig. 8:** Sum of the horizontal and vertical $\beta$-functions as a function on the phase advance $\varphi$

Large beam dimensions need large apertures of the quadrupole and dipole magnets in the ring. Running the machine at the highest energy can therefore lead to limitations on the focusing power, as the gradient of a quadrupole lens scales as the inverse of its squared aperture radius, i.e., $g \propto 1/r^2$, and this increases the cost of the magnet lenses. Therefore it is recommended not to tune the lattice too far away from the ideal phase advance.

Here, for completeness, I have to make a short remark about electron machines. Unlike the situation in proton rings, electron beams are flat in general: owing to the damping mechanism of synchrotron radiation [4], the vertical emittance of an electron or positron beam is only a small fraction of the horizontal emittance, such that $\varepsilon_y \approx 1–10\%\ \varepsilon_x$. The calculation for the optimization of the phase advance can and should be restricted to the horizontal plane only, and the condition for the smallest beam dimension is

$$\frac{d}{d\varphi}(\hat{\beta}) = \frac{d}{d\varphi}\frac{(1+\sin(\varphi/2))L}{\sin\varphi} = 0 \quad \Rightarrow \varphi = 76°.$$

Figure 9 shows the horizontal and vertical values of $\beta$ as a function of $\varphi$ in this case. In an electron ring, the typical phase advance $\varphi$ is in the range of 60–90°.

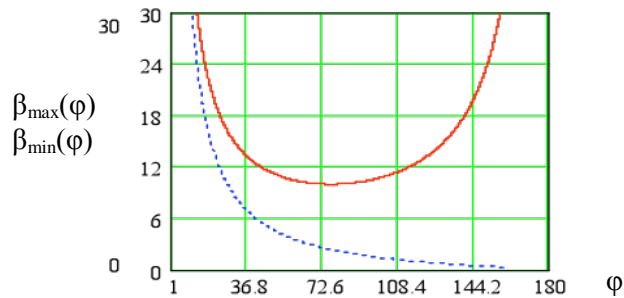

**Fig. 9:** Horizontal and vertical $\beta$ in a FODO cell as a function of the phase advance $\varphi$

## 2.6  Dispersion in a FODO lattice

In our treatment of the design of a magnet lattice and our description of the optical parameters, we have restricted ourselves until now to the case where all particles have the ideal momentum. In doing so, we have followed the usual path in all books about linear beam dynamics. But, in general, the energy (or momentum) of the particles stored in a ring follows a certain distribution and deviates from the ideal momentum $p_0$ of the beam.

We know from linear beam optics [1] that the differential equation for the transverse motion gains an additional term if the momentum deviation is not zero: $\Delta p/p \neq 0$. We obtain an inhomogeneous equation of motion

$$x'' + K(s) \cdot x = \frac{1}{\rho} \frac{\Delta p}{p}. \tag{17}$$

The left-hand side of Eq. (17) is the same as that in the homogeneous Eq. (2), and the parameter $K$ describes the focusing strength of the lattice element at the position $s$ in the ring. As usual, the general solution of Eq. (17) is the sum of the complete solution $x_h$ of the homogeneous equation and a special solution of the inhomogeneous equation, $x_i$:

$$x_h'' + K(s) x_h = 0,$$
$$x_i'' + K(s) x_i = \frac{1}{\rho} \cdot \frac{\Delta p}{p}.$$

The special solution $x_i$ can be normalized to the momentum error $\Delta p/p$, and we obtain the so-called dispersion function $D(s)$:

$$x_i(s) = D(s) \cdot \frac{\Delta p}{p}. \tag{18}$$

This describes the additional amplitude of the particle oscillation due to the momentum error and is created by the $1/\rho$ term, i.e., in general, by the bending fields of the dipole magnets of our storage ring.

Starting as before from initial conditions $x_0$ and $x_0'$, the general solution for the particle trajectory now reads

$$x(s) = C(s) x_0 + S(s) x_0' + D(s) \frac{\Delta p}{p}$$

or, including the expression for the angle $x'(s)$,

$$\begin{pmatrix} x \\ x' \end{pmatrix}_s = \begin{pmatrix} C & S \\ C' & S' \end{pmatrix} \cdot \begin{pmatrix} x \\ x' \end{pmatrix}_0 + \frac{\Delta p}{p} \begin{pmatrix} D \\ D' \end{pmatrix}.$$

For convenience, in general the matrix is extended to include the second term and written

$$\begin{pmatrix} x \\ x' \\ \Delta p/p \end{pmatrix}_s = \begin{pmatrix} C & S & D \\ C' & S' & D' \\ 0 & 0 & 1 \end{pmatrix} \cdot \begin{pmatrix} x \\ x' \\ \Delta p/p \end{pmatrix}_0.$$

The dispersion function $D(s)$ is (obviously) defined by the focusing properties of the lattice and the bending strength of the dipole magnets $1/\rho$, and it can be shown that [1]

$$D(s) = S(s) \cdot \int_{s_0}^{s} \frac{1}{\rho(\tilde{s})} C(\tilde{s}) \, d\tilde{s} - C(s) \cdot \int_{s_0}^{s} \frac{1}{\rho(\tilde{s})} S(\tilde{s}) \, d\tilde{s}. \tag{19}$$

The variable $s$ refers to the position where the dispersion is obtained (or measured, if you like), and the integration has to be performed over all places $\tilde{s}$ where a non-vanishing term $1/\rho$ exists (in general, in the dipole magnets of the ring).

### 2.6.1 *Example*

The 2 × 2 matrix for a drift space is given by

$$M_{\text{Drift}} = \begin{pmatrix} C & S \\ C' & S' \end{pmatrix} = \begin{pmatrix} 1 & \ell \\ 0 & 1 \end{pmatrix}.$$

As there are no dipoles in the drift space, the $1/\rho$ term in Eq. (19) is zero and we obtain the extended 3 × 3 matrix

$$M_{\text{Drift}} = \begin{pmatrix} 1 & \ell & 0 \\ 0 & 1 & 0 \\ 0 & 0 & 1 \end{pmatrix}.$$

To calculate the dispersion in a FODO cell, including the $1/\rho$ term of the dipoles, things look quite different: We refer again to the thin-lens approximation that has already been used for the calculation of the $\beta$-functions. The matrix for a half cell has been derived above (see Eq. (10)). Again, we want to point out that in the thin-lens approximation, the length $\ell$ of the drift space is just half the length of the cell, as the quadrupole lenses have zero length. The matrix for a half cell is

$$M_{\text{halfcell}} = \begin{pmatrix} C & S \\ C' & S' \end{pmatrix} = \begin{pmatrix} 1-\dfrac{\ell}{\tilde{f}} & \ell \\ \dfrac{-\ell}{\tilde{f}^2} & 1+\dfrac{\ell}{\tilde{f}} \end{pmatrix}.$$

Using this expression, we can calculate the terms $D$, $D'$ of the 3 × 3 matrix:

$$D(s) = S(s) \cdot \int_{s0}^{s} \frac{1}{\rho(\tilde{s})} C(\tilde{s})\, d\tilde{s} - C(s) \cdot \int_{s0}^{s} \frac{1}{\rho(\tilde{s})} S(\tilde{s})\, d\tilde{s},$$

$$D(\ell) = \frac{\ell}{\rho}\left(\ell - \frac{\ell^2}{2\tilde{f}}\right) - \left(1 - \frac{\ell}{\tilde{f}}\right)\frac{1}{\rho}\frac{\ell^2}{2} = \frac{\ell^2}{\rho} - \frac{\ell^3}{2\tilde{f}\rho} - \frac{\ell^2}{2\rho} + \frac{\ell^3}{2\tilde{f}\rho},$$

$$D(\ell) = \frac{\ell^2}{2\rho}.$$

In an analogous way, we can derive an expression for $D'$,

$$D'(\ell) = \frac{\ell}{\rho}\left(1 + \frac{\ell}{2\tilde{f}}\right),$$

and we obtain the complete matrix for a FODO half cell,

$$M_{\text{half cell}} = \begin{pmatrix} C & S & D \\ C' & S' & D' \\ 0 & 0 & 1 \end{pmatrix} = \begin{pmatrix} 1-\frac{\ell}{\tilde{f}} & \ell & \frac{\ell^2}{2\rho} \\ \frac{-\ell}{\tilde{f}^2} & 1+\frac{\ell}{\tilde{f}} & \frac{\ell}{\rho}\left(1+\frac{\ell}{2\tilde{f}}\right) \\ 0 & 0 & 1 \end{pmatrix}.$$

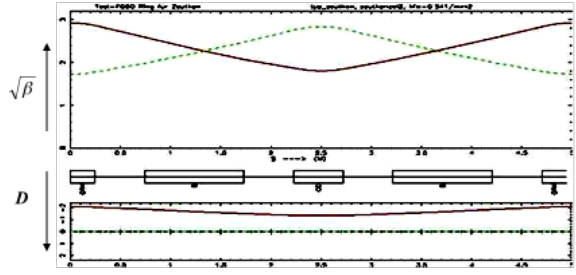

**Fig. 10:** $\beta$-function (top) and horizontal dispersion (bottom) in a FODO cell

Now we know that, owing to symmetry, the dispersion in a FODO lattice reaches its maximum value in the centre of a focusing quadrupole and its minimum in a defocusing quadrupole, as shown in Fig. 10, for example, where, in addition to the $\beta$-function, the dispersion is shown in the lower part of the figure. Therefore we obtain the boundary conditions for the transformation from a focusing to a defocusing lens,

$$\begin{pmatrix} \check{D} \\ 0 \\ 1 \end{pmatrix} = M_{1/2} \begin{pmatrix} \hat{D} \\ 0 \\ 1 \end{pmatrix},$$

which can be used to calculate the dispersion at these locations:

$$\check{D} = \hat{D}\left(1-\frac{\ell}{\tilde{f}}\right)+\frac{\ell^2}{2\rho},$$

$$0 = \frac{-\ell}{\tilde{f}^2}\hat{D} + \frac{\ell}{\rho}\left(1+\frac{\ell}{2\tilde{f}}\right).$$

Remember that we have to use the focal length of a half quadrupole,

$$\tilde{f} = 2f,$$

and that the phase advance is given by

$$\sin(\varphi/2) = \frac{L_{\text{cell}}}{2\tilde{f}}.$$

We obtain the following expressions for the maximum dispersion in the centre of a focusing quadrupole and for the minimum dispersion in the centre of a defocusing lens:

$$\hat{D} = \frac{\ell^2}{\rho}\frac{(1+(1/2)\sin(\frac{\phi_{\text{cell}}}{2}))}{\sin^2(\frac{\phi_{\text{cell}}}{2})},$$

$$\check{D} = \frac{\ell^2}{\rho}\frac{(1-(1/2)\sin(\frac{\phi_{\text{cell}}}{2}))}{\sin^2(\frac{\phi_{\text{cell}}}{2})},$$

(20)

It is interesting to note that the dispersion depends only on the half length $\ell$ of the cell, the bending strength of the dipole magnet $1/\rho$, and the phase advance $\varphi$. The dependence of $D$ on the phase advance is shown in the plot in Fig. 11. The two values $D_{\max}$ and $D_{\min}$ decrease with increasing phase $\varphi$ (which is just another way of saying 'for increasing focusing strength', as $\varphi$ depends on the focusing strength of the quadrupole magnets).

To summarize these considerations, I would like to make the following remarks:

- A small dispersion needs strong focusing and therefore a large phase advance.
- There is, however, an optimum phase advance concerning the best (i.e., smallest) value of the $\beta$-function.
- Furthermore, the stability criterion limits the choice of the phase advance per cell.

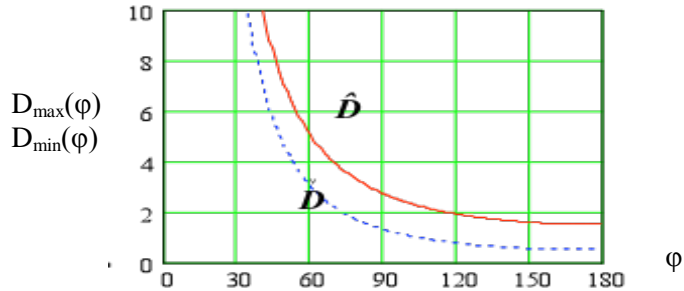

**Fig. 11:** Dispersion at the focusing and defocusing quadrupole lenses in a FODO as a function of the phase $\varphi$

In general, therefore, one has to find a compromise for the focusing strength in a lattice that takes into account the stability of the motion, the $\beta$-functions in both transverse planes, and the dispersion function. In a typical high-energy machine this optimization is not too difficult, as the dispersion does not have too much impact on the beam parameters (as long as it is compensated at the interaction point of the two beams).

In synchrotron light sources, however, the beam emittance is usually the parameter that has to be optimized (this means, in nearly all cases, *minimized*), and as the emittance depends on the dispersion $D$ in an electron storage ring, the dispersion function and its optimization are of the greatest importance in these machines. In an electron ring, the horizontal beam emittance is given by the expression

$$\varepsilon_x = \frac{55}{32\sqrt{3}} \frac{\hbar}{mc} \gamma^2 \frac{\left\langle \frac{1}{R^3} H(s) \right\rangle}{J_x \left\langle \frac{1}{R^2} \right\rangle},$$

where the function $H(s)$ is defined by

$$H(s) = \gamma D^2 + 2\alpha DD' + \beta D'^2.$$

The optimization of $H(s)$ in a magnet structure is a subject of its own, and an introduction to the field of the so-called low-emittance lattices can be found, for example, in [5].

## 2.7 Orbit distortions in a periodic lattice

The lattice that we have designed so far consists only of a small number of basic elements: bending magnets that define the geometry of the circular accelerator and, for a given particle momentum, the size of the machine; and quadrupole lenses that define the phase advance of the single-particle trajectories and, through this parameter, define the beam dimensions and the stability of the motion. Now it is time to fill the empty spaces in the lattice cell with some useful other components, which means we have to talk about the 'O's of the FODO.

Nobody is perfect, and this statement also holds for storage rings. In the case of a dipole magnet, an error in the bending field can be described by an additional kick $\theta$ (typically measured in mrad) on the particles,

$$\theta = \frac{ds}{\rho} = \frac{\int Bds}{p/e}$$

The beam oscillates in the corresponding plane, and the resulting amplitude of the orbit is

$$x(s) = \frac{\sqrt{\beta(s)}}{2\sin(\pi Q)} \oint \beta(\tilde{s}) \frac{1}{\rho(\tilde{s})} \cos(|\varphi(\tilde{s}) - \varphi(s)| - \pi Q) d\tilde{s}. \tag{21}$$

This is given by the $\beta$-function at the place of the dipole magnet $\beta(\tilde{s})$ and its bending strength $1/\rho$, and the $\beta$-function at the observation point in the lattice $\beta(s)$. For the lattice designer, this means that if a correction magnet has to be installed in the lattice cell, it should be placed at a location where $\beta$ is high in the corresponding plane.

At the same time, Eq. (21) tells us that the amplitude of an orbit distortion is highest at a place where $\beta$ is high, and this is the place where beam position monitors have to be located to measure the orbit distortion precisely. In practice, therefore, both the beam position monitors and the orbit correction coils are located at places in the lattice cell where the $\beta$-function in the plane considered is large, i.e., close to the corresponding quadrupole lens.

## 2.8 Chromaticity in a FODO cell

The chromaticity $Q'$ describes an optical error of a quadrupole lens in an accelerator. For a given magnetic field, i.e., gradient of the quadrupole magnet, particles with smaller momentum will feel a stronger focusing force.

The chromaticity $Q'$ relates the resulting tune shift to the relative momentum error of the particle:

$$\Delta Q = Q' \cdot \frac{\Delta p}{p}.$$

As it is a consequence of the focusing properties of the quadrupole magnets, it is given by the characteristics of the lattice. For small momentum errors $\Delta p/p$, the focusing parameter $k$ can be written as

$$k(p) = \frac{g}{p/e} = g \cdot \frac{e}{p_0 + \Delta p},$$

where $g$ denotes the gradient of the quadrupole lens and $p_0$ the design momentum, and the term $\Delta p$ refers to the momentum error. If $\Delta p$ is small, as we have assumed, we can write

$$k(p) \approx g \cdot \frac{e}{p_0}(1 - \frac{\Delta p}{p}) = k + \Delta k.$$

This describes a quadrupole error

$$\Delta k = -k_0 \cdot \frac{\Delta p}{p}$$

and leads to a tune shift of

$$\Delta Q = \frac{1}{4\pi} \int \Delta k \cdot \beta(s) ds,$$

or

$$\Delta Q = \frac{-1}{4\pi} \frac{\Delta p}{p} \int k_0 \cdot \beta(s) ds.$$

By definition, the chromaticity $Q'$ of a lattice is therefore given by

$$Q' = \frac{-1}{4\pi} \int \beta(s) k(s) ds . \tag{22}$$

Let us assume now that the accelerator consists of $N$ identical FODO cells. Then, replacing $\beta(s)$ by its maximum value at the focusing quadrupoles and by its minimum value at the defocusing quadrupoles, we can approximate the integral by a sum:

$$Q' = \frac{-1}{4\pi} N \frac{\hat{\beta} - \check{\beta}}{f_Q},$$

where $f_Q = 1/(k*\ell)$ denotes the focal length of the quadrupole magnet. We obtain

$$Q' = \frac{-1}{4\pi} N \frac{1}{f_Q} \left\{ \frac{L(1+\sin(\varphi/2)) - L(1-\sin(\varphi/2))}{\sin\varphi} \right\}. \tag{23}$$

Here, we have used Eq. (15) for $\check{\beta}$ and $\hat{\beta}$. With some useful trigonometric transformations such as

$$\sin x = 2\sin\frac{x}{2}\cos\frac{x}{2},$$

we can transform the right-hand side of Eq. (23) to obtain

$$Q' = \frac{-1}{4\pi} N \frac{1}{f_Q} \left\{ \frac{L \cdot \sin(\varphi/2)}{\sin(\varphi/2)\cos(\varphi/2)} \right\},$$

or, for one single cell ($N = 1$),

$$Q' = \frac{-1}{4\pi} \frac{1}{f_Q} \left\{ \frac{L \cdot \tan(\varphi/2)}{\sin(\varphi/2)} \right\}.$$

Remembering the relation

$$\sin\frac{\varphi}{2} = \frac{L}{4f_Q},$$

we obtain a surprisingly simple result for the chromaticity contribution of a FODO cell,

$$Q' = \frac{-1}{4\pi} \tan(\varphi/2) .$$

## 3 Lattice insertions

We have seen in Fig. 2 that the lattice of a typical machine for the acceleration of high-energy particles consists of two quite different types of parts: the arcs, which are built from a number of identical cells, and the straight sections that connect them and that house complicated systems such as dispersion suppressors, mini-beta insertions, and high-energy particle detectors.

### 3.1 Drift space

To provide some initial insight into the design of lattice insertions, I would like to start with some comments concerning a simple drift space embedded in a normal lattice structure.

What would happen to the beam parameters $\alpha$, $\beta$, and $\gamma$ if we were to stop focusing for a while? The transfer matrix for the Twiss parameters from the position 0 to the position $s$ in a lattice is given by the formula

$$\begin{pmatrix} \beta \\ \alpha \\ \gamma \end{pmatrix}_s = \begin{pmatrix} C^2 & -2SC & S^2 \\ -CC' & SC'+S'C & -SS' \\ C'^2 & -S'C' & S'^2 \end{pmatrix} \cdot \begin{pmatrix} \beta \\ \alpha \\ \gamma \end{pmatrix}_0, \qquad (24)$$

where the cosine and sine functions $C$ and $S$ are given by the focusing properties of the lattice elements between the two points (Eqs. (3) and (7)). For a drift space of length $s$, this is, according to Eq. (3), as simple as

$$M = \begin{pmatrix} C & S \\ C' & S' \end{pmatrix} = \begin{pmatrix} 1 & s \\ 0 & 1 \end{pmatrix},$$

and the optical parameters will develop as a function of $s$ in the following way:

$$\begin{aligned} \beta(s) &= \beta_0 - 2\alpha s + \gamma_0 s^2, \\ \alpha(s) &= \alpha_0 - \gamma_0 s, \\ \gamma(s) &= \gamma_0. \end{aligned} \qquad (25)$$

We shall now take a closer look at these relations.

#### 3.1.1 Location of the beam waist

From the first of these equations, we see immediately that if the drift space is long enough, even a convergent beam at the position 0 will become divergent, as the term $\gamma_0 s^2$ is always positive. This is shown schematically in Fig. 12.

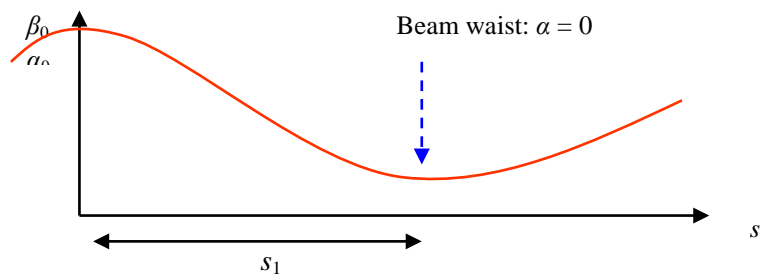

**Fig. 12:** Schematic drawing of a beam waist

Therefore there will be a point in the drift space where the beam dimension is smallest, in other words where the beam envelope has a waist. The position of this waist can be calculated by requiring

$$\alpha(s_1) = 0,$$

and the second equation of Eq. (25) then gives

$$\alpha_0 = \gamma_0 \cdot s_1,$$

or

$$s_1 = \frac{\alpha_0}{\gamma_0}.$$

The position of the waist is given by the ratio of the $\alpha$- and $\gamma$-functions at the beginning of the drift space.

As the parameter $\gamma$ is constant in a drift space and $\alpha$ is zero at the waist, we can directly calculate the beam size that we obtain at the waist:

$$\gamma(s_1) = \gamma_0, \quad \alpha(s_1) = 0,$$

$$\beta(s_1) = \frac{1 + \alpha^2(s_1)}{\gamma(s_1)} = \frac{1}{\gamma_0}. \tag{26}$$

The $\beta$-function at the location of the waist is given by the inverse of the $\gamma$-function at the beginning of the drift space: a nice and simple scaling law.

### 3.1.2  *β-function in a drift space*

It is worth thinking a little bit more about the behaviour of the Twiss parameters in a drift space. Namely, the scaling of $\beta$ with the length of the drift space has a large impact on the design of a lattice. At specific locations in the ring, we have to create places for beam instrumentation, for beam injection and extraction, and for the installation of particle detectors. Let us assume that we are in the centre of a drift space and that the situation is symmetric, which means that the index *0* refers to the position at the starting point, but now we want to have left–right symmetric optics with respect to it, so that $\alpha_0 = 0$. From Eq. (25), we obtain at the starting point

$$\beta(s) = \beta_0 - 2\alpha_0 s + \gamma_0 s^2,$$

and knowing already from Eq. (26) that $\alpha = 0$ at the waist, we have

$$\gamma_0 = \frac{1 + \alpha_0^2}{\beta_0} = \frac{1}{\beta_0}.$$

Hence we obtain $\beta$ as a function of the distance *s* from the starting point:

$$\beta(s) = \beta_0 + \frac{s^2}{\beta_0}. \tag{27}$$

I would like to point out two facts in this context:

- Equation (27) is a direct consequence of Liouville's theorem: the density of the phase space of the particles is constant in an accelerator. In other words, if there are only conservative forces, the beam emittance $\varepsilon$ is constant, which leads immediately to Eq. (24) and finally Eq. (27). And, as the conservation of $\varepsilon$ is a fundamental law, there is no trick that can be used to avoid it and no way to overcome the increase of the beam dimension in a drift space.

- The behaviour of β in a drift space has a strong impact on the design of a storage ring. As large beam dimensions have to be avoided, this means that large drift spaces are forbidden or at least very inconvenient. We shall see in the next section that this places one of the major limitations on the luminosity of colliding beams in an accelerator.

At the beam waist, we can derive another short relation that is often used for the scaling of beam parameters. The beam envelope $\sigma$ is given by the β-function and the emittance of the beam by

$$\sigma(s) = \sqrt{\varepsilon \cdot \beta(s)},$$

and the divergence $\sigma'$ is given by

$$\sigma'(s) = \sqrt{\varepsilon \cdot \gamma(s)}.$$

Now, as $\gamma = (1 + \alpha^2)/\beta$, wherever $\alpha = 0$ the beam envelope has a local minimum (i.e., a waist) or maximum. At that position, the β-function is just the ratio of the beam envelope and the beam divergence:

$$\beta(s) = \frac{\sigma(s)}{\sigma'(s)} \quad \text{at a waist.}$$

If we cannot fight against Liouville's theorem, we can at least try to optimize its consequences. Equation (27) for β in a symmetric drift space can be used to find the starting value that gives the smallest beam dimension at the end of a drift space of length $\ell$. Setting

$$\frac{d\hat{\beta}}{d\beta_0} = 1 - \frac{\ell^2}{\beta_0^2} = 0$$

gives us the value of $\beta_0$ that leads to the smallest β after a drift space of length $\ell$:

$$\beta_0 = \ell. \tag{28}$$

For a starting value of $\beta_0 = \ell$ at the waist, the maximum beam dimension at the end of the drift space is smallest, and its value is just double the length of the drift space:

$$\hat{\beta} = 2\beta_0 = 2\ell.$$

## 3.2 Mini-beta insertions and luminosity

The discussion in the previous section has shown that the β-function in a drift space can be chosen with respect to the length $\ell$ to minimize the beam dimensions and, according to those dimensions, the aperture requirements for vacuum chambers and magnets. In general, the value of β is of the order of some metres and the typical length of the drift spaces in a lattice is of the same order.

However, the straight sections of a storage ring are often designed for the collision of two counter-rotating beams, and the β-functions at the collision points are therefore very small compared with their values in the arc cells. Typical values are more in the range of *centimetres* than of *metres*. Nevertheless, the same scaling law (28) holds, and the optimum length $\ell$ of such a drift space would be, for example, approximately 36 cm for the interaction regions of the two beams in the HERA collider. Modern high-energy detectors, in contrast, are impressive devices that consist of many large components, and they do not fit into a drift space of a few centimetres. Figure 13 shows, as an example, the ZEUS detector at the HERA collider. It is evident that a special treatment of the storage ring lattice is needed for the installation of such a huge detector.

The lattice therefore has to be modified before and after the interaction point, to establish a large drift space in which the detector for the high-energy experiment can be embedded. At the same time, the beams have to be focused strongly to obtain very small beam dimensions in both transverse planes at the collision point or, in other words, to obtain high luminosity. Such a lattice structure is called a 'mini-beta insertion'.

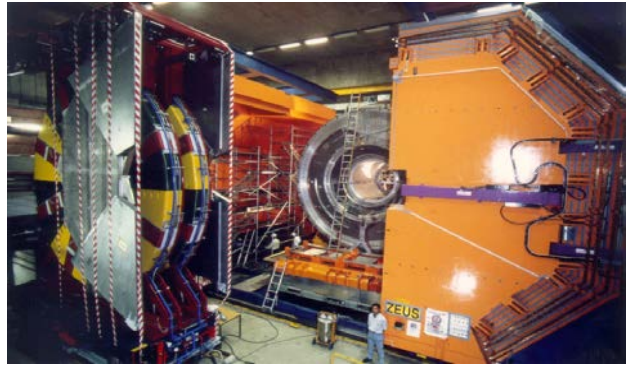

**Fig. 13:** Particle detector for the ZEUS collaboration at the HERA storage ring

The luminosity of a particle collider is defined by the event rate $R$ of a specific reaction (e.g., the production of a particle in the collision of the beams):

$$R = \sigma_R \cdot L$$

The production rate of a reaction is given by its physics cross-section $\sigma_R$ and a number that is the result of the lattice design: the luminosity $L$ of the storage ring. This is determined by the beam optics at the collision point and the magnitudes of the stored beam currents [6]:

$$L = \frac{1}{4\pi e^2 f_0 b} \cdot \frac{I_1 \cdot I_2}{\sigma_x^* \cdot \sigma_y^*}$$

Here $I_1$ and $I_2$ are the values of the stored beam currents, $f_0$ is the revolution frequency of the machine, and $b$ is the number of stored bunches. The quantities $\sigma_x^*$ and $\sigma_y^*$ in the denominator are the beam sizes in the horizontal and vertical planes at the interaction point. For a high-luminosity collider, the stored beam currents have to be high and, at the same time, the beams have to be focused at the interaction point to very small dimensions.

Figure 14 shows the typical layout of such a mini-beta insertion. It consists in general of

– a symmetric drift space that is large enough to house the particle detector and whose beam waist (where $\alpha_0 = 0$) is centred at the interaction point of the colliding beams;

– a quadrupole doublet (or triplet) on each side as close as possible;

– additional quadrupole lenses to match the Twiss parameters of the mini-beta insertion to the optical parameters of the lattice cell in the arc.

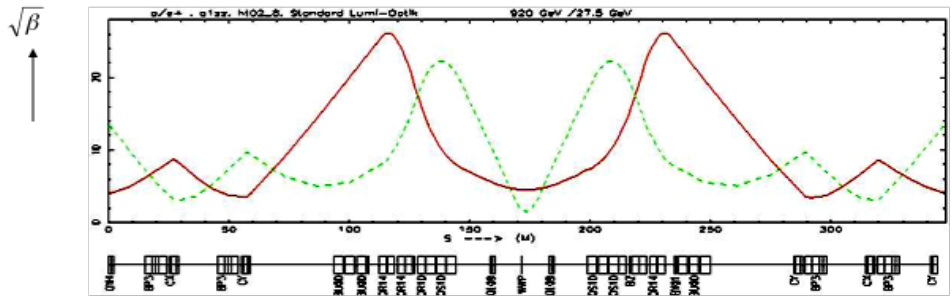

**Fig. 14:** Layout of a mini-beta insertion scheme

As a mini-beta scheme is always a kind of symmetric drift space, we can apply the formula that we have derived above. For $\alpha = 0$, we obtain a quadratic increase of the $\beta$-function in the drift space, and at the distance $\ell_1$ of the first quadrupole lens we obtain

$$\beta(s) = \beta_0 + \frac{\ell_1^2}{\beta_0}.$$

The size of the beam at the position of the second quadrupole can be calculated in a similar way. As shown in Fig. 15, the transfer matrix of the quadrupole doublet system consists of four parts, namely two drift spaces with lengths $\ell_1$ and $\ell_2$, and a focusing and a defocusing quadrupole magnet. Starting at the injection point (IP), we obtain, again in the thin-lens approximation,

$$M_{D1} = \begin{pmatrix} 1 & \ell_1 \\ 0 & 1 \end{pmatrix}, \qquad M_{f1} = \begin{pmatrix} 1 & 0 \\ \frac{1}{f_1} & 1 \end{pmatrix},$$

$$M_{D2} = \begin{pmatrix} 1 & \ell_2 \\ 0 & 1 \end{pmatrix}, \qquad M_{f2} = \begin{pmatrix} 1 & 0 \\ \frac{-1}{f_2} & 1 \end{pmatrix}.$$

It should be noted that in general, the first lens of such a system is focusing in the vertical plane and therefore, according to the sign convention used in this school, the focal length is positive, i.e., $1/f_1 > 0$. The matrix for the complete system is

$$M = M_{QF} \cdot M_{D2} \cdot M_{QD} \cdot M_{D1}$$

$$M = \begin{pmatrix} 1 & 0 \\ -1/f_2 & 1 \end{pmatrix} \cdot \begin{pmatrix} 1 & \ell_2 \\ 0 & 1 \end{pmatrix} \cdot \begin{pmatrix} 1 & 0 \\ 1/f_1 & 1 \end{pmatrix} \cdot \begin{pmatrix} 1 & \ell_1 \\ 0 & 1 \end{pmatrix}.$$

Multiplying out, we obtain

$$M = \begin{pmatrix} 1 + \frac{\ell_2}{f_1} & \ell_1 + \ell_2 + \frac{\ell_1 \ell_2}{f_1} \\ \frac{1}{f_1} - \frac{1}{f_2} - \frac{\ell_2}{f_1 f_2} & -\frac{\ell_1}{f_2} - \frac{\ell_1 \ell_2}{f_1 f_2} - \frac{\ell_2}{f_2} + \frac{\ell_1}{f_1} + 1 \end{pmatrix} = \begin{pmatrix} C & S \\ C' & S' \end{pmatrix}.$$

Remembering the transformation of the Twiss parameters in terms of matrix elements (see Eq. (6))

$$\begin{pmatrix} \beta \\ \alpha \\ \gamma \end{pmatrix}_s = \begin{pmatrix} C^2 & -2SC & S^2 \\ -CC' & SC' + S'C & -SS' \\ C'^2 & -S'C' & S'^2 \end{pmatrix} \cdot \begin{pmatrix} \beta \\ \alpha \\ \gamma \end{pmatrix}_0,$$

we put in the terms from above and obtain

$$\beta(s) = C^2 \beta_0 - 2SC \alpha_0 + S^2 \gamma_0.$$

Here, the index 0 denotes the interaction point and $s$ refers to the position of the second quadrupole lens. As we are starting at the IP, where $\alpha_0 = 0$ and $\gamma_0 = 1/\beta_0$, we can simplify this equation and obtain

$$\beta(s) = C^2\beta_0 + S^2/\beta_0,$$

$$\beta(s) = \beta_0 * \left(1 + \frac{\ell_2}{f_1}\right)^2 + \frac{1}{\beta_0} * \left(\ell_1 + \ell_2 + \frac{\ell_1\ell_2}{f_1}\right)^2.$$

This formula for $\beta$ at the second quadrupole lens is very useful when the gradient and aperture of a mini-beta quadrupole magnet have to be designed.

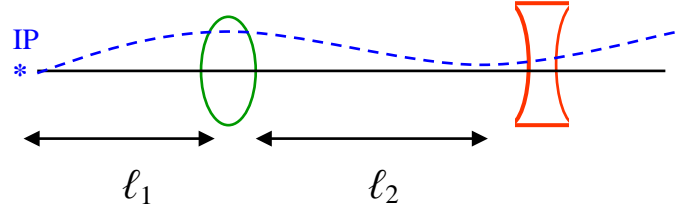

**Fig. 15:** Schematic layout of a mini-beta doublet

### *3.2.1 Phase advance in a mini-beta insertion*

Unlike the situation in an arc, where the phase advance is a function of the focusing properties of the cell, in a mini-beta insertion or in any long drift space it is quasi-constant. As we know from linear beam optics, the phase advance is given by

$$\phi(s) = \int \frac{1}{\beta(s)} ds,$$

and, inserting $\beta(s)$ from Eq. (27), we obtain

$$\phi(s) = \frac{1}{\beta_0} \int_0^{\ell_1} \frac{1}{1 + s^2/\beta_0^2} ds ,$$

$$\phi(s) = \arctan \frac{\ell_1}{\beta_0} ,$$

where $\ell_1$ denotes the distance of the first focusing element from the IP, i.e., the length of the first drift space. In Fig. 16, the phase advance is plotted as a function of $\ell$ for a $\beta$-function of 10 cm. If the length of the drift space is large compared with the value of $\beta$ at the IP, which is usually the case, the phase advance is approximately 90° on each side. In other words, the tune of the accelerator increases by half an integer in the complete drift space.

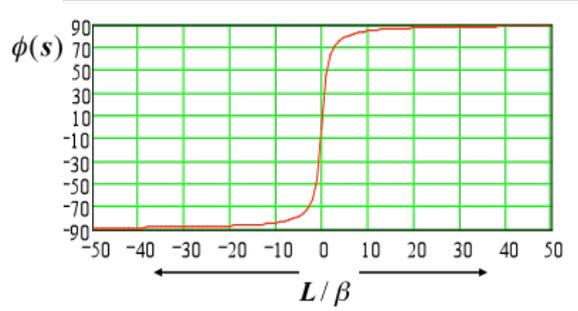

**Fig. 16:** Phase advance in a symmetric drift space as a function of the drift length

There are some further points that can be made concerning mini-beta sections, which will not be discussed here in detail but only mentioned briefly. As we have seen, large values of the β-function on either side of the interaction point cannot be avoided if a mini-beta section is inserted into a machine lattice. These high β values have a strong impact on the machine performance:

– According to Eq. (22), the chromaticity of a lattice is given by the strength of the focusing elements and the value of the β-function at that position:

$$Q' = \frac{-1}{4\pi} \int \beta(s) k(s) \, \mathrm{d}s \; .$$

In a mini-beta insertion, unfortunately, we have both strong quadrupoles and large beam dimensions. The contribution of such a lattice section to $Q'$ can therefore be very large and, as it has to be corrected in the ring, it is in general a strong limitation on the luminosity in a collider ring.

– As the beam dimensions in an insertion can reach large values, the aperture of the mini-beta magnets has to be much larger than in the FODO structure of an arc. Large magnet apertures, however, limit the strength of the quadrupole. Here, a compromise has to be found between the aperture requirements, the integrated focusing strength, the spot size at the IP, and cost, as large magnets are quite expensive.

– Last but not least, the problem of field quality and adjustment has to be mentioned. Compared with the standard magnets in an arc, the lenses in a mini-beta section have to fulfil stronger requirements. A kick due to a dipole error or to an off-centre quadrupole lens leads to an orbit distortion that is proportional to the β-function at the place of the error (Eq. (21)).

As a consequence the field quality concerning higher multipole components has to be much higher and the adjustment of the mini-beta quadrupoles much more precise than for the quadrupole lenses in an arc. In general, multipole components of the order of $\Delta B/B = 10^{-4}$ with respect to the main field and alignment tolerances in the transverse plane of about a tenth of a millimetre are desired.

### 3.2.2 *Guidelines for the design of a mini-beta insertion*

– First, calculate the periodic solution for a lattice cell in the arc. This will serve to provide starting values for the insertion.

– Introduce the drift space needed for the insertion device (e.g., a particle detector).

– Put the mini-beta quadrupoles as close as possible to the IP — often, nowadays, these lenses are embedded in the detector to keep the distance *s* small.

– Introduce additional quadrupole lenses to match the optical parameters of the insertion to the solution for the arc cell. In general, the functions $\alpha_x$, $\beta_x$, $\alpha_y$, $\beta_y$ and the horizontal dispersion

$D_x$, $D'_x$ have to be matched. Sometimes additional quadrupoles are needed to adjust the phase advance in both planes, and in the case of HERA even the vertical dispersion $D_y$, $D'_y$ needs to be corrected. So, at least eight additional magnet lenses are required.

# 4  Dispersion suppressors

The dispersion function $D(s)$ has already been mentioned in Section 2.6, where we have shown that it is a function of the focusing and bending properties of the lattice cell, and we calculated its size as a function of the cell parameters.

Now we have to return to this topic in the context of lattice insertions. In the interaction region of an accelerator, which means the straight section of the ring where two counter-rotating beams collide (typically designed as a mini-beta insertion), the dispersion function $D(s)$ has to vanish. A non-vanishing dispersion dilutes the luminosity of the machine and leads to additional stop bands in the working diagram of the accelerator (synchro betatron resonances) that are driven by the beam–beam interaction. Therefore, sections have to be inserted in our magnet lattice that are designed to reduce the function $D(s)$ to zero, called dispersion-suppressing schemes. In Eq. (18), we have shown that the oscillation amplitude of a particle is given by

$$x(s) = x_\beta(s) + D(s) \cdot \frac{\Delta p}{p}.$$

Here, $x_\beta$ describes the solution of the homogeneous differential equation (which is valid for particles with the ideal momentum $p_0$), and the second term—the dispersion term—describes the additional oscillation amplitude for particles with a relative momentum error $\Delta p/p_0$.

As an example, let me present some numbers for the HERA proton storage ring. The beam size at the collision point of the two beams in the horizontal and vertical directions is determined by the mini-beta insertion, and has values $\sigma_x \approx 118$ μm and $\sigma_y \approx 32$ μm. The contribution $x_D$ of the dispersion function to the oscillation amplitude of the particles, for a typical dispersion in the cell of $D(s) \approx 1.5$ m and a momentum distribution of the beam $\Delta p/p \approx 5 \times 10^{-4}$, is equal to 0.75 mm. Therefore a mini-beta insertion in general has to be combined with a lattice sructure to suppress the dispersion at the IP.

## 4.1  Dispersion suppression using additional quadrupole magnets

### 4.1.1  The 'straightforward way'

There are several ways to suppress the dispersion, and each of them has its advantages and disadvantages. We shall not present all of them here; instead, we shall restrict ourselves to the basic idea behind dispersion suppression. We will assume that a periodic lattice is given and that we simply want to continue the FODO structure of the arc through the straight section—but with vanishing dispersion. Given an optical solution in the arc cells, as shown for example in Fig. 17, we have to guarantee that, starting from the periodic solution for the optical parameters $\alpha(s)$, $\beta(s)$, and $D(s)$, we obtain a situation at the end of the suppressor where we have $D(s) = D'(s) = 0$ and the values for $\alpha$ and $\beta$ are unchanged.

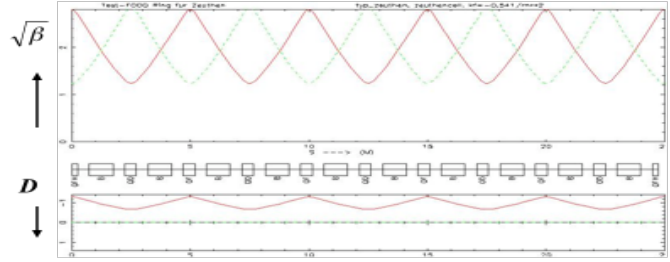

**Fig. 17:** Periodic FODO, including the horizontal dispersion function in the lower part of the plot

The boundary conditions

$$D(s) = D'(s) = 0,$$
$$\beta_x(s) = \beta_{x\,\text{arc}}, \quad \alpha_x(s) = \alpha_{x\,\text{arc}},$$
$$\beta_y(s) = \beta_{y\,\text{arc}}, \quad \alpha_y(s) = \alpha_{y\,\text{arc}}$$

can be fulfilled by introducing six additional quadrupole lenses, whose strengths have to be matched individually in a suitable way. This can be done by using one of the beam optics codes that are available today in every accelerator laboratory. An example is shown in Fig. 18, starting from a FODO structure with a phase advance of $\varphi \approx 61°$ per cell. The advantages of this scheme are:

- it works for an arbitrary phase advance of the arc structure;
- matching also works for different optical parameters $\alpha$ and $\beta$ before and after the dispersion suppressor;
- the ring geometry is unchanged, as no additional dipoles are needed.

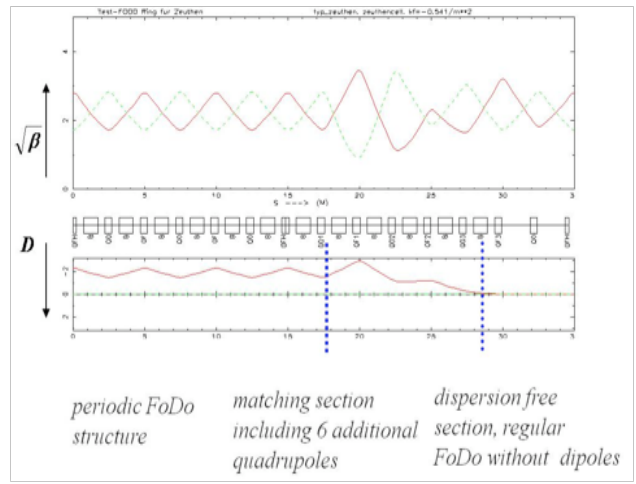

**Fig. 18:** Dispersion suppressor using individually powered quadrupole lenses

On the other hand, there are a number of disadvantages that have to be mentioned:

- as the strengths of the additional quadrupole magnets have to be matched individually, the scheme needs additional power supplies and quadrupole magnet types, which can be an expensive requirement;
- the required quadrupole fields are, in general, stronger than in the arc;

- the *β*-function reaches higher values (sometimes *really* high values), and so the aperture of the vacuum chamber and of the magnets has to be increased.

However, there are alternative ways to suppress the dispersion that do not need individually powered quadrupole lenses but instead change the strength of the dipole magnets at the end of the arc structure.

### 4.1.2  The 'clever way': Half-bend schemes

These dispersion-suppressing schemes consist of *n* additional FODO cells that are added to the periodic arc structure but where the bending strength of the dipole magnets is reduced. As before, we split the lattice into three parts: the periodic structure of the FODO cells in the arc, the lattice insertion in which the dispersion is suppressed, and a following dispersion-free section, which may be another FODO structure without bending magnets, a mini-beta insertion, or something else.

The calculation of the suppressor proceeds in several steps.

*Step 1: establish the matrix for a periodic arc cell.* We have already calculated the dispersion in a FODO lattice (see Eq. (20)), where we derived a formula for *D* in the thin-lens approximation as a function of the focusing properties of the lattice. Now we have to be a little more accurate and, instead of using the focusing strength and phase advance, we have to work with the optical parameters of the system. We know that the transfer matrix for the lattice of a storage ring can be written as a function of the optical parameters in Eq. (14):

$$M_{0 \to s} = \begin{pmatrix} \sqrt{\dfrac{\beta_s}{\beta_0}}(\cos\phi + \alpha_0 \sin\phi) & \sqrt{\beta_s \beta_0}\sin\phi \\ \dfrac{(\alpha_0 - \alpha_s)\cos\phi - (1+\alpha_0\alpha_s)\sin\phi}{\sqrt{\beta_s \beta_0}} & \sqrt{\dfrac{\beta_0}{\beta_s}}(\cos\phi - \alpha_s \sin\phi) \end{pmatrix}. \quad (29)$$

The variable $\phi$ refers to the phase advance between the starting point *0* and the end point *s* of the transformation. This formula is valid for any starting and end points in the lattice. If, for convenience, we refer the transformation to the centre of a focusing quadrupole magnet (as we have usually done in the past), where $\alpha = 0$, and if we are interested in the solution for a complete cell, we can write the equation in a simpler form. Extending the matrix to the $3 \times 3$ form to include the dispersion terms (see Section 2.6) and taking the periodicity of the system into account, so that $\beta_0 = \beta_s$, we obtain

$$M_{\text{cell}} = \begin{pmatrix} C & S & D \\ C' & S' & D' \\ 0 & 0 & 1 \end{pmatrix} = \begin{pmatrix} \cos\phi_c & \beta_c \sin\phi_c & D(l) \\ \dfrac{-1}{\beta_c}\sin\phi_c & \cos\phi_c & D'(l) \\ 0 & 0 & 1 \end{pmatrix}. \quad (30)$$

Now $\phi_c$ is the phase advance for a single cell, and the index 'c' reminds us that we are talking about the periodic solution (one complete *cell*).

The dispersion elements *D* and *D'* are, as usual, given by the elements *C* and *S* according to Eq. (19):

$$D(\ell) = S(\ell) \int_0^\ell \frac{1}{\rho(\tilde{s})} C(\tilde{s})d\tilde{s} - C(\ell)\int_0^\ell \frac{1}{\rho(\tilde{s})}S(\tilde{s})d\tilde{s},$$

$$D'(\ell) = S'(\ell) \int_0^\ell \frac{1}{\rho(\tilde{s})} C(\tilde{s})d\tilde{s} - C'(\ell)\int_0^\ell \frac{1}{\rho(\tilde{s})}S(\tilde{s})d\tilde{s}.$$

The values $C(\ell)$ and $S(\ell)$ refer to the symmetry point of the cell (the centre of the quadrupole). The integrals, however, have to be taken over the dipole magnet, where $\rho \neq 0$. Assuming a constant bending radius in the dipole magnets of the arc, i.e., $\rho$ = const (which is a good approximation in general), we can evaluate the integrals over $C(s)$ and $S(s)$ if we approximate their values by those in the centre of the dipole magnet.

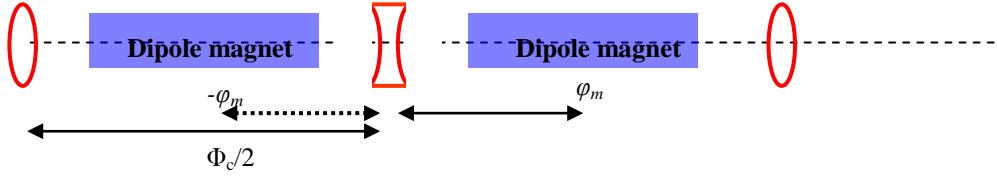

**Fig. 19:** Schematic view of a FODO: notation for the phase relations in the cell

*Step 2: transformation of the optical functions from the centre of the quadrupole to the centre of the dipole, to calculate the functions $C(\tilde{s})$ and $S(\tilde{s})$.* As indicated in the schematic layout in Fig. 19, we have to transform the optical functions $\alpha$ and $\beta$ from the centre of the quadrupole lens to the centre of the dipole magnet. The formalism is given by Eq. (29), and we obtain (with $\alpha_0 = 0$)

$$C_m = \sqrt{\frac{\beta_m}{\beta_c}} \cos \Delta\phi = \sqrt{\frac{\beta_m}{\beta_c}} \cos\left(\frac{\phi_c}{2} \pm \varphi_m\right),$$

$$S_m = \sqrt{\beta_m \beta_c} \sin\left(\frac{\phi_c}{2} \pm \varphi_m\right).$$

The index 'm' tells us that we are dealing with values in the centre of the bending magnets now, and as our starting point was the centre of the focusing quadrupole, the phase advance for this transformation is half the phase advance of the cell that brings us to the defocusing lens, plus or minus the phase distance $\varphi_m$ from that point to the centre of the dipole.

Now we can evaluate the integrals for $D(s)$ and $D'(s)$:

$$D(\ell) = \beta_c \sin\phi_c \frac{L_B}{\rho} \sqrt{\frac{\beta_m}{\beta_c}} \cos\left(\frac{\phi_c}{2} \pm \varphi_m\right) - \cos\phi_c \frac{L_B}{\rho} \sqrt{\beta_m \beta_c} \sin\left(\frac{\phi_c}{2} \pm \varphi_m\right), \tag{31}$$

where $L_B$ is the length of the dipole magnets. Putting $\delta = L_B/\rho$ for the bending angle, we obtain

$$D(\ell) = \delta \sqrt{\beta_m \beta_c} \left\{\sin\phi_c \left[\cos\left(\frac{\phi_c}{2} + \varphi_m\right) + \cos\left(\frac{\phi_c}{2} - \varphi_m\right)\right] - \cos\phi_c \left[\sin\left(\frac{\phi_c}{2} + \varphi_m\right) + \sin\left(\frac{\phi_c}{2} - \varphi_m\right)\right]\right\}.$$

Using the trigonometric relations

$$\cos x + \cos y = 2 \cdot \cos\frac{x+y}{2} \cdot \cos\frac{x-y}{2},$$

$$\sin x + \sin y = 2 \cdot \sin\frac{x+y}{2} \cdot \cos\frac{x-y}{2},$$

we obtain

$$D(\ell) = \delta \sqrt{\beta_m \beta_c} \left\{\sin\phi_c \, 2\cos\frac{\phi_c}{2} \cos\varphi_m - \cos\phi_c \, 2\sin\frac{\phi_c}{2} \cos\varphi_m\right\},$$

$$D(\ell) = 2\delta \sqrt{\beta_m \beta_c} \cos\varphi_m \left\{\sin\phi_c \cos\frac{\phi_c}{2} - \cos\phi_c \sin\frac{\phi_c}{2}\right\},$$

and with

$$\sin 2x = 2\sin x \cdot \cos x,$$
$$\cos 2x = \cos^2 x - \sin^2 x,$$

we can derive the dispersion at the centre of the quadrupole magnet in its final form:

$$D(\ell) = 2\delta\sqrt{\beta_m \beta_c} \cos\varphi_m \left\{2\sin\frac{\phi_c}{2}\cos^2\frac{\phi_c}{2} - \left(\cos^2\frac{\phi_c}{2} - \sin^2\frac{\phi_c}{2}\right)\sin\frac{\phi_c}{2}\right\},$$
$$D(\ell) = 2\delta\sqrt{\beta_m \beta_c} \cos\varphi_m \sin\frac{\phi_c}{2}\left\{2\cos^2\frac{\phi_c}{2} - \cos^2\frac{\phi_c}{2} + \sin^2\frac{\phi_c}{2}\right\},$$
$$D(\ell) = 2\delta\sqrt{\beta_m \beta_c} \cos\varphi_m \sin\frac{\phi_c}{2}. \tag{32}$$

This is the expression for the dispersion term of the matrix in Eq. (30) at the centre of the quadrupole magnet, determined from the dipole strength $1/\rho$ and matrix elements $C$ and $S$ at the position of the dipole.

In full analogy, we can derive a formula for the derivative of the dispersion, $D'(s)$:

$$D'(\ell) = 2\delta\sqrt{\beta_m/\beta_c}\cos\varphi_m \cos\frac{\phi_c}{2}. \tag{33}$$

As we are referring to the situation in the centre of a quadrupole, the expressions for $D(s)$ and $D'(s)$ are valid for a periodic structure, namely one FODO cell. Therefore we require periodic boundary conditions for the transformation from one cell to the next:

$$\begin{pmatrix} D_c \\ D'_c \\ 1 \end{pmatrix} = M_c \cdot \begin{pmatrix} D_c \\ D'_c \\ 1 \end{pmatrix}$$

and, by symmetry,

$$D'_c = 0. \tag{34}$$

With these boundary conditions, the periodic dispersion in the FODO cell is determined:

$$D_c = D_c \cdot \cos\phi_c + \delta\sqrt{\beta_m \beta_c} \cdot \cos\varphi_m \cdot 2\sin\frac{\phi_c}{2},$$
$$D_c = \delta\sqrt{\beta_m \beta_c} \cdot \cos\varphi_m / \sin\frac{\phi_c}{2}. \tag{35}$$

*Step 3: Calculate the dispersion in the suppressor part.* In the dispersion suppressor section, starting with the value at the end of the cell, $D(s)$ is reduced to zero. Or, turning the problem around and thinking from right to left, the dispersion has to be created, starting from $D = D' = 0$. The goal is to generate the dispersion in this section in such a way that the values of the periodic arc cell are obtained.

The relation for $D(s)$ still holds in the same way:

$$D(\ell) = S(\ell)\int_0^\ell \frac{1}{\rho(\tilde{s})}C(\tilde{s})d\tilde{s} - C(\ell)\int_0^\ell \frac{1}{\rho(\tilde{s})}S(\tilde{s})d\tilde{s}.$$

But we can now take several cells into account (the number of cells in the suppressor scheme), and we have the freedom to choose a dipole strength $\rho_{\text{suppr}}$ in this section that differs from the strength of the arc dipoles. As the dispersion is generated in $n$ cells, the matrix for these $n$ cells is

$$M_n = M_c^n = \begin{pmatrix} \cos n\phi_c & \beta_c \sin n\phi_c & D_n \\ \dfrac{-1}{\beta_c}\sin n\phi_c & \cos n\phi_c & D'_n \\ 0 & 0 & 1 \end{pmatrix}$$

and, according to Eq. (31), the dispersion created in these $n$ cells is given by

$$D_n = \beta_c \sin n\phi_c \cdot \delta_{\text{sup}} \cdot \sum_{i=1}^{n} \cos\left(i\phi_c - \frac{1}{2}\phi_c \pm \varphi_m\right) \cdot \sqrt{\frac{\beta_m}{\beta_c}} - \cos n\phi_c \cdot \delta_{\text{sup}} \cdot \sum_{i=1}^{n} \sqrt{\beta_m \beta_c} \cdot \sin\left(i\phi_c - \frac{1}{2}\phi_c \pm \varphi_m\right)$$

$$D_n = \sqrt{\beta_m \beta_c} \cdot \sin n\phi_c \cdot \delta_{\text{sup}} \sum_{i=1}^{n} \cos\left((2i-1)\frac{\phi_c}{2} \pm \varphi_m\right) - \sqrt{\beta_m \beta_c} \cos n\phi_c \cdot \delta_{\text{sup}} \sum_{i=1}^{n} \sin\left((2i-1)\frac{\phi_c}{2} \pm \varphi_m\right)$$

Remembering the trigonometric gymnastics shown above, we obtain

$$D_n = \delta_{\text{sup}} \cdot \sqrt{\beta_m \beta_c} \cdot \sin n\phi_c \sum_{i=1}^{n} \cos\left((2i-1)\frac{\phi_c}{2}\right) \cdot 2\cos\varphi_m$$

$$-\delta_{\text{sup}} \cdot \sqrt{\beta_m \beta_c} \cdot \cos n\phi_c \sum_{i=1}^{n} \sin\left((2i-1)\frac{\phi_c}{2}\right) \cdot 2\cos\varphi_m$$

$$D_n = 2\delta_{\text{sup}} \cdot \sqrt{\beta_m \beta_c} \cdot \cos\varphi_m \left\{ \sum_{i=1}^{n} \cos\left((2i-1)\frac{\phi_c}{2}\right) \cdot \sin n\phi_c - \sum_{i=1}^{n} \sin\left((2i-1)\frac{\phi_c}{2}\right) \cdot \cos n\phi_c \right\}$$

$$D_n = 2\delta_{\text{sup}} \sqrt{\beta_m \beta_c} \cdot \cos\varphi_m \sin(n\phi_c) \cdot \frac{\sin(n\phi_c/2) \cdot \cos(n\phi_c/2)}{\sin(\phi_c/2)}$$

$$-2\delta_{\text{sup}} \sqrt{\beta_m \beta_c} \cdot \cos\varphi_m \cos(n\phi_c) \frac{\sin(n\phi_c/2) \cdot \sin(n\phi_c/2)}{\sin(\phi_c/2)}$$

$$D_n = \frac{2\delta_{\text{sup}} \sqrt{\beta_m \beta_c} \cdot \cos\varphi_m}{\sin(\phi_c/2)} \left\{ 2\sin\frac{n\phi_c}{2}\cos\frac{n\phi_c}{2} \cdot \cos\frac{n\phi_c}{2}\sin\frac{n\phi_c}{2} - \left(\cos^2\frac{n\phi_c}{2} - \sin^2\frac{n\phi_c}{2}\right)\sin^2\frac{n\phi_c}{2} \right\}$$

And, finally,

$$D_n = \frac{2\delta_{\text{sup}} \sqrt{\beta_m \beta_c} \cdot \cos\varphi_m}{\sin(\phi_c/2)} \sin^2\frac{n\phi_c}{2}. \tag{36}$$

This relation gives us the dispersion $D(s)$ that is created in $n$ cells that have a phase advance of $\Phi_c$ per cell; $\delta_{\text{sup}}$ is the bending strength of the dipole magnets located in these $n$ cells; and the optical functions $\beta_m$ and $\beta_c$ refer to the values at the centre of the dipole and of the quadrupole, respectively.

In a similar calculation, we obtain an expression for the derivative $D'(s)$ of the dispersion:

$$D'_n = \frac{2\delta_{sup}\sqrt{\beta_m/\beta_c} \cdot \cos\varphi_m}{\sin(\phi_c/2)} \sin n\phi_c. \tag{37}$$

*Step 4: Determine the strength of the suppressor dipoles.* The last step is to calculate the strength of the dipole magnets in the suppressor section. As the dispersion generated in this section has to be equal to that of the arc cells for the optimum match of $D$, we equate the expressions (34), (35) and (36), (37). For $D_n$, we obtain the condition

$$D_n = \frac{2\delta_{sup}\sqrt{\beta_m\beta_c} \cdot \cos\varphi_m}{\sin(\phi_c/2)} \sin^2 \frac{n\phi_c}{2} = \frac{\delta_{arc}\sqrt{\beta_m\beta_c} \cdot \cos\varphi_m}{\sin(\phi_c/2)}$$

and, for $D'$,

$$D'_n = \frac{2\delta_{sup}\sqrt{\beta_m/\beta_c} \cdot \cos\varphi_m}{\sin(\phi_c/2)} \sin n\phi_c = 0.$$

From the latter two equations, we deduce two conditions for the dispersion matching:

$$\left. \begin{array}{l} 2\delta_{suppr} \sin^2\left(\dfrac{n\phi_c}{2}\right) = \delta_{arc} \\ \sin(n\phi_c) = 0 \end{array} \right\} \quad \delta_{suppr} = \frac{1}{2}\delta_{arc}. \tag{38}$$

If the phase advance per cell in the arc fulfils the condition $\sin(n\phi_c) = 0$, the strength of the dipoles in the suppressor region is just half the strength of the arc dipoles. In other words, the phase has to be chosen as

$$n\phi_c = k \cdot \pi, \qquad k = 1, 3, ...$$

There are a number of possible phase advances that fulfil this relation, but clearly not every arbitrary phase is allowed. Two possible combinations are $\phi_c = 90°$, $n = 2$ cells and $\phi_c = 60°$, $n = 3$ cells in the suppressor.

Figure 20 shows such a half-bend dispersion suppressor, starting from a FODO structure with a 60° phase advance per cell. The focusing strengths of the FODO cells before and after the suppressor are identical, with the exception that — clearly — the FODO cells on the right are 'empty', i.e., they have no bending magnets. It is evident that unlike the case for the suppressor scheme based on quadrupole lenses, the $\beta$-function is now unchanged in the suppressor region.

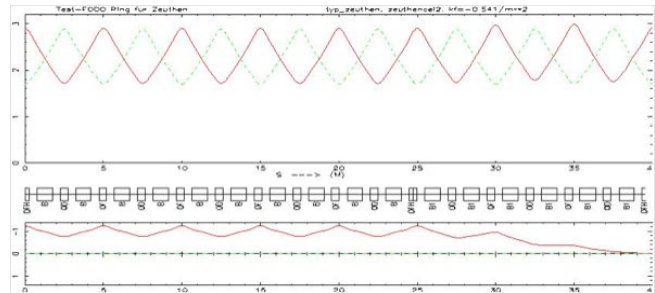

**Fig. 20:** Dispersion suppressor based on a half-bend scheme

Again, this scheme has advantages:

– no additional quadrupole lenses are needed, and no individual power supplies;
– the aperture requirements are just the same as those in the arc, as the $\beta$-functions are unchanged;

and disadvantages:

– it works only for certain values of the phase advance in the structure and therefore restricts the freedom of choice for the optics in the arc;
– special dipole magnets are needed (having half the strength of the arc type);
– the geometry of the ring is changed.

I should mention here, for purists only, that in these equations the phase advance of the suppressor part is equal to that of the arc structure — but this is not completely true, as the weak-focusing term $1/\rho^2$ in the arc FODO differs from the term $1/(2\rho)^2$ in the half-bend scheme. As, however, the impact of the weak focusing on the beam optics can be neglected in many practical cases, Eq. (38) is *nearly* correct.

The application of such a scheme is very elegant, but as it has a strong impact on the beam optics and geometry, it has to be embedded in the accelerator design at an early stage.

## 4.2 The missing-bend dispersion suppressor scheme

For completeness, I would like to present another suppressor scheme, which is also used in a number of storage rings. This consists of $n$ cells without dipole magnets at the end of an arc, followed by $m$ cells that are identical to the arc cells. The matching condition for this 'missing-bend scheme' with respect to the phase advance is

$$\frac{2m+n}{2}\Phi_C = (2k+1)\frac{\pi}{2},$$

and the condition for the number $m$ of cells required is

$$\sin\frac{m\phi_c}{2} = \frac{1}{2}, \quad k = 0, 2, \ldots \quad \text{or} \quad \sin\frac{m\phi_c}{2} = \frac{-1}{2}, \quad k = 1, 3, \ldots$$

An example based on $\Phi = 60°$ and $m = n = 1$ is shown in Fig. 21. A variety of similar scenarios is feasible for different phase relations in the arc and the corresponding bending strength needed to reduce $D(s)$ [7, 8].

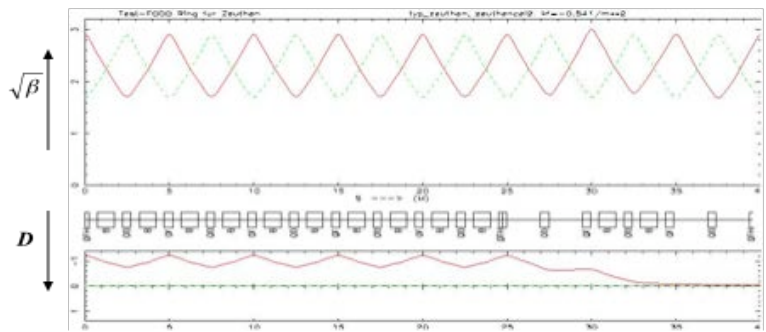

**Fig. 21:** Dispersion suppressor based on a missing-magnet scheme

In general, one of the above two schemes (missing-bend or half-bend suppressor) is combined with a number of individual quadrupole lenses to guarantee the flexibility of the system with respect to phase changes in the lattice and to keep the size of the $\beta$-function moderate.